\documentclass[10pt,journal,twocolumn]{IEEEtran}
 \usepackage{graphicx}
 \usepackage{amsmath}
 \usepackage{cases}
  \usepackage{amsfonts}
\usepackage{indentfirst}
\usepackage{caption}
\usepackage{cite}
\usepackage{algorithm}
\usepackage{algpseudocode}
\usepackage{booktabs}
\usepackage{subfigure}
\usepackage{multirow}
\usepackage{amsthm}

 \theoremstyle{remark}

\theoremstyle{}
\newtheorem{theorem}{Theorem}
\theoremstyle{}

\newtheorem{lemma}{Lemma}
\theoremstyle{}

\theoremstyle{remark}

\title{On the Gap Between Decentralized and Centralized Coded Caching Schemes}

\author{Qifa~Yan,~
        Xiaohu~Tang,~
        and~Qingchun~Chen
\thanks{The authors are with The School of Information Science and Technology, Southwest Jiaotong University, Chengdu, China, 610031. E-mails: qifa@my.swjtu.edu.cn,  xhutang@swjtu.edu.cn, qcchen@swjtu.edu.cn.}
}
\begin{document}
\maketitle

\begin{abstract}
 Caching is a promising solution to satisfy the ongoing explosive demands for  multi-media traffics. Recently, Maddah-Ali and Niesen  proposed both centralized and decentralized coded caching schemes, which are able to attain significant performance gains over  uncoded caching schemes. Particular, their work indicates that there exists a performance gap between the decentralized coded caching scheme and the centralized coded caching scheme. In this paper, we investigate this gap. As a result,  we  prove that the multiplicative gap (\emph{i.e.}, the ratio of their performances) is between $1$ and $1.5$.  The upper bound tightens the original one of $12$ by Maddah-Ali and Niesen, while  the lower bound verifies the intuition that the centralized coded caching scheme always outperforms its decentralized counterpart. Notably, both bounds are  achievable  in some cases.  Furthermore, we prove that  the gap can be arbitrarily close to $1$ if the number of users is large enough, which suggests the great potential in practical applications to use the less optimal but more practical decentralized coded caching scheme.
\end{abstract}

\IEEEpeerreviewmaketitle

\section{Introduction}\label{sec_introduction}
With the dramatic increasing demands for video streams in many applications such as Youtube and Netflix, a huge burden is placed on the underlying networks that deliver the streaming data to users. One promising technique to mitigate this burden is  caching \cite{caire2013femtocaching}, whereby relatively popular contents are prefetched at the local cache memories of the end users. Based on the cached contents and users’ demands, the server sends a signal to enable the users decode their requested
files. With the help of the content in the caches, the amount of the transmission is reduced.

Usually, the caching network operates in two distinct phases: In the \emph{placement phase}, some fractions of the content are disseminated to all the users'   caches in off peak times. It should be noted that the operations in placement phase do not depend on the actual user requests, since the server does not have that information at this moment. In the \emph{delivery phase} that happens in peak time, the user requests are uncovered.  Then the server responds by sending a signal over the shared link to satisfy the user requests. In conventional caching design, the key idea  is to deliver part of the content to caches close to the end users, so that the requested content can be served locally. Coding in the cached content or/and the transmitted signal was typically not considered. Recently, in their seminal work  \cite{maddah2013fundamental,maddah2013decentralized}, Maddah-Ali and Niesen showed that, by exploiting the cached content in the users, multicasting opportunities can be created and thus the central server is able to send distinct content to different users by jointly designing the cached contents and delivered signals. This achieves a global gain depending on the normalized cumulative caching size in reducing the amount of delivered signals.   This type of technique is termed \emph{coded caching}.

The proposed coded caching schemes  in \cite{maddah2013fundamental} and \cite{maddah2013decentralized} are essentially  \emph{centralized}  and \emph{decentralized} respectively.
 In the centralized coded caching scheme \cite{maddah2013fundamental,yan2015placement}, the placement phase should be coordinated by a central sever carefully so that the content in different caches  overlap to create coded-multicasting opportunities among users with different demands. In the  delivery phase, once receiving the requests of users, the server sends a predesignated signal that fully exploits the coded-multicasting opportunities. Whereas, in the decentralized coded caching scheme \cite{maddah2013decentralized}, 
 the cached contents  at the placement phase are independent across the users. Later in the delivery phase, after acquiring the information about the set of users, their cached contents
and requests, the server sends a signal that efficiently exploits the coded-multicasting opportunities existed in the contents of the end users.

Obviously, an advantage of the decentralized coded caching scheme is that it extends the application of coded caching. In real networks, the centralized coordination may not be available at the placement phase in many cases. For example, the server may not have the knowledge of  the identity or even just the number of active users in advance, or the users may be in distinct networks during the placement phases. Whereas, the decentralized coded caching scheme can be applied to such circumstances,  such as online scenarios \cite{maddah2013online} and random access \cite{Yan2017Online}. Notably, Maddah-Ali and Niesen proved in \cite{maddah2013decentralized} that, the performance of decentralized coded caching scheme is  within a constant multiplicative gap  of its centralized counterpart, \emph{i.e.}, the ratio of their performances does not exceed $12$. Moreover,  they claimed  that the gap can be tighten to $1.6$ numerically for any system parameters.

   Notably, the scheme proposed by Maddah-Ali and Niesen achieves the best performance within uncoded scheme in most cases \cite{wan2016uncoded}. Thus, it is meaningful to investigate the aforementioned performance gap. Specifically, we prove that the gap is between $1$ and $1.5$ for any system parameter setup theoretically. The upper bound tightens the proved bound $12$ and justifies the claimed numerical bound $1.6$ by  Maddah-Ali and Niesen. The lower bound verifies that  the decentralized scheme is always inferior to its centralized counterpart. Though the later conclusion seems intuitively correct,  the theoretical justification has not been found to the best knowledge of the authors. Furthermore, we show that in case of large number of users, the gap  always approaches $1$. This indicates that in large systems, the decentralized coded caching scheme can be utilized in terms of both the achieved performance and realization requirement.

  The remainder of this paper is organized as follows: In Section \ref{sec_start}, we review the network model in \cite{maddah2013fundamental,maddah2013decentralized}  and the corresponding schemes and performances.  In Section \ref{sec_results}, we present the main result. Section \ref{sec_proof}  gives the proof of the main result. Finally, we conclude this paper in Section \ref{sec_conclusion}.

\section{Network Model and Coded Caching Schemes}\label{sec_start}

We review the model studied by Maddah-Ali and Niesen in \cite{maddah2013fundamental,maddah2013decentralized} firstly, and then present their schemes and results in this section.

\subsection{Network Model}
\begin{figure}[htbp]
\centering\includegraphics[width=0.45\textwidth]{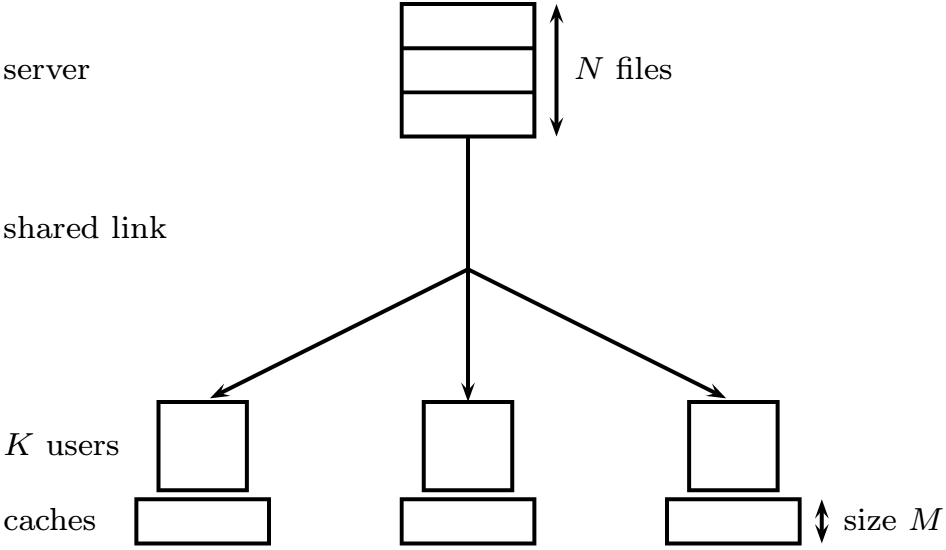}
\caption{A central server with $N$ popular files connected to $K$ users through an error-free shared link. Each file is of $F$ bits and each user is equipped with a cache of size $MF$ bits. In the figure, $N=K=3$, $M=1$.}\label{fig_system}
\end{figure}

Consider a  system consisting of a  server with $N$ popular files $W_1,W_2,\cdots, W_N$, each  of size $F$ bits. The server is connected to $K$ ($K\geq2$) users  $\mathcal{K}$ through an error-free  shared  link,   where $\mathcal{K}=\{1,2,\cdots,K\}$ is the set of users.  Each user $k$ is provisioned with an isolated  cache $Z_k$ with size $MF$ bits, where $M\in(0,N]$\footnote{For notational simplicity, we consider $M\in(0,N]$ and $K\geq 2$. It is  trivial to include the case $M=0$ or/and $K=1$ in the result. }. The system is illustrated in Fig.  \ref{fig_system}.

The system operates in two  phases:

\begin{enumerate}
  \item[$1.$]  In the placement phase, the users are able to access  the whole files $W_1,W_2,\cdots,W_N$ at the server. Then each user $k$ fills its cache $Z_k$ with some content relevant to the files. The only constraint is that, the size of the content should not exceed its cache size $MF$ bits.
  \item[$2.$] In the delivery phase, each user $k$ requests  $W_{d_k}$ from the server  where $k\in
\mathcal{K}$ and $d_k\in\{1,2,\cdots,N\}$. Then the server responds by sending a signal  of $R^{(d_1,\cdots,d_K)}F$ bits to the users through the shared link. Then each user $k$ decodes its requested file $W_{d_k}$ from the received signal  as well as the pre-stored content in its cache $Z_k$. Specifically,  $R^{(d_1,\cdots,d_K)}$ is called the rate of the system under the request $(d_1,\cdots,d_K)$.
\end{enumerate}

Generally speaking, given the parameters $M,N,K$ and the request $(d_1,\cdots,d_K)$,   $R^{(d_1,\cdots,d_K)}$ should be as small as possible.
Let
\begin{align}
R\overset{\triangle}{=}\max_{d_1,\cdots,d_K} R^{(d_1,\cdots,d_K)}\notag
\end{align}
be the worst-case normalized rate for  a caching scheme. Therefore, the objective  is to design schemes to minimize $R$ in the literature.

\subsection{Maddah-Ali and Niesen's Coded Caching Schemes}

Maddah-Ali and Niesen first investigated the above model and proposed a centralized and decentralized coded caching scheme in \cite{maddah2013fundamental} and \cite{maddah2013decentralized}, respectively. Both schemes are proved to outperform the conventional uncoded scheme, whose rate is given by \footnote{To simplify the notations, the symbols $N,K$ do not appear in notations of achievable rates in \eqref{eq_RU}, \eqref{eq_RC} and \eqref{eq_RD}. }
\begin{eqnarray}\label{Eqn_nocode}
R_{U}(M)=K\left(1-\frac{M}{N}\right)\cdot\min\left\{1,\frac{N}{K}\right\}\label{eq_RU}
\end{eqnarray}

  In the centralized scheme, the central server is able to jointly design the content placement in the caches $Z_1,Z_2,\cdots,Z_K$. While in a decentralized coded caching scheme, the content in caches $Z_1,Z_2,\cdots,Z_K$ should be independent of each other.  As  mentioned in Section \ref{sec_introduction}, since the decentralized scheme is easier to be implemented compared to the centralized scheme, it is worth clarifying the performance difference between them.


Theorem \ref{thmC} elaborates the performance of the centralized scheme in \cite{maddah2013fundamental}.

\begin{theorem}\label{thmC}(Centralized Coded Caching \cite{maddah2013fundamental}) For $N\in\mathbb{N}^+$ files and $K\in\mathbb{N}^+$ users each with cache of size $M=s\cdot N/K$, $s\in\{0,1,\cdots,K\}$, the centralized coded caching scheme achieves a rate
\begin{eqnarray}
R_C(M)&\overset{\triangle}{=}& K\cdot\left(1-\frac{M}{N}\right)\cdot\min\left\{\frac{1}{1+KM/N},\frac{N}{K}\right\}\nonumber\\
&=&(K-s)\cdot\min\left\{\frac{1}{1+s},\frac{N}{K}\right\}\label{eq_RC}
\end{eqnarray}
For general $0< M\leq N$, the lower convex envelop of these points is achievable, i.e.,
\begin{eqnarray}\label{eq_RM}
R_C(M)=\theta R_C((s-1)N/K)+(1-\theta) R_C(sN/K)
\end{eqnarray}
where $s\in\{1,2,\cdots,K\}$  and $\theta\in [0,1)$ satisfy
\begin{align}
M=\theta\cdot(s-1)\frac{N}{K}+(1-\theta)\cdot s \frac{N}{K}\notag
\end{align}
\end{theorem}

\vspace{3mm}

For $M\in\{0,N/K,2N/K,\cdots,N\}$, Algorithm \ref{alg1} illustrates the coded caching scheme proposed in \cite{maddah2013fundamental}, where $\oplus$ in Line \ref{oplus} is  bit-wise XOR operation.  It was shown in \cite{maddah2013fundamental} that Algorithm \ref{alg1} achieves the following rate
\begin{eqnarray}\label{Eqn_center}
K\left(1-\frac{M}{N}\right)\frac{1}{1+KM/N}
\end{eqnarray}
Comparing \eqref{Eqn_nocode} and \eqref{Eqn_center} with \eqref{eq_RC}, we see that the rate in \eqref{eq_RC} is a result of combining Algorithm \ref{alg1} and the conventional uncoded caching scheme (\emph{i.e.,} the server always chooses the approach that leads to smaller rate between Algorithm \ref{alg1} and the conventional uncoded scheme). For more general $M\in[0,N]$,  any point on the lower convex envelop given by \eqref{eq_RM}  is also achievable by employing so-called memory sharing technique \cite{maddah2013fundamental}.

\begin{algorithm}[htb]
\caption{Centralized Coded Caching Scheme}\label{alg1}
\begin{algorithmic}[1]
\Procedure {Placement}{$W_1,\cdots,W_{N}$}
\State $t\leftarrow\frac{KM}{N}$
\State $\mathfrak{T}\leftarrow\{\mathcal{T}\subset \mathcal{K}:|\mathcal{T}|=t\}$
\For{$n\in\{1,\cdots,N\}$}
\State Split $W_n$ into $\{W_{n,\mathcal{T}}:\mathcal{T}\in\mathfrak{T}\}$ of equal size
\EndFor
\For{$k\in\mathcal{K}$}
\State $Z_k\leftarrow\{W_{n,\mathcal{T}}:n\in\{1,\cdots,N\},\mathcal{T}\in\mathfrak{T},k\in\mathcal{T}\}$
\EndFor
\EndProcedure

\quad

\Procedure{Delivery}{$W_1,\cdots,W_{N},d_1,\cdots,d_{K}$}
\State  $t\leftarrow\frac{KM}{N}$
\State $\mathfrak{S}\leftarrow\{\mathcal{S}\subset\mathcal{K}:|\mathcal{S}|=t+1\}$
\State Server sends $\{\oplus_{k\in\mathcal{S}} W_{d_k,\mathcal{S}\backslash \{k\}}:\mathcal{S}\in\mathfrak{S}\}$\label{oplus}
\EndProcedure
\end{algorithmic}
\end{algorithm}

Theorem \ref{thmD} depicts the performance of the decentralized coded caching scheme in \cite{maddah2013decentralized}.

\begin{theorem}\label{thmD}(Decentralized Coded Caching \cite{maddah2013decentralized}) For $N\in\mathbb{N}^+$ files and $K\in\mathbb{N}^+$ users each with cache of size $M\in(0,N]$, the decentralized coded caching scheme  achieves the following rate
\begin{align}
&R_D(M)\overset{\triangle}{=}\notag\\
&K\cdot\left(1-\frac{M}{N}\right)\cdot\min\left\{\frac{N}{KM}\left(1-\left(1-\frac{M}{N}\right)^K\right),\frac{N}{K}\right\}\label{eq_RD}
\end{align}
\end{theorem}

The rate in \eqref{eq_RD} can be achieved by employing Algorithm \ref{alg2},   where $V_{k,\mathcal{S}}$ in Line \ref{Vs} corresponds the bits of file $W_{d_k}$ that are present in the cache of every user in $\mathcal{S}$ and absent in the cache of the users outside $\mathcal{S}$, and all the elements $V_{k,\mathcal{S}\backslash\{k\}}$ are assumed to be zero padded to the length of the longest element. Algorithm \ref{alg2} composes of two delivery procedures, where Delivery1  and Delivery2  correspond to the first and second items in the braces of \eqref{eq_RD} respectively. In fact, it  selects the procedure with smaller rate.
\begin{algorithm}[htb]
\caption{Decentralized Coded Caching Scheme}\label{alg2}
\begin{algorithmic}[1]
\Procedure {Placement}{$W_1,W_2,\cdots,W_{N}$}
\For{$k\in\mathcal{K},n\in\{1,\cdots,N\}$}
\State User $k$ independently caches a subset of $MF/N$ bits of file $W_n$, chosen uniformly at random
\EndFor
\EndProcedure

\quad

\Procedure{Delivery1}{$W_1,\cdots,W_{N},d_1,\cdots,d_{K}$}
\For {$s=K,K-1,\cdots,1$}
\For{ $\mathcal{S}\subset\mathcal{K}: |\mathcal{S}|=s$}
\State Server sends $\oplus_{k\in\mathcal{S}}V_{k,\mathcal{S}\backslash \{k\}}$\label{Vs}
\EndFor
\EndFor
\EndProcedure

\quad

\Procedure{Delivery2}{$W_1,\cdots,W_{N},d_1,\cdots,d_{K}$}
\For {$n\in\{1,\cdots,N\}$}
\State Server sends enough random linear combinations of bits in file $W_n$ for all users requesting it to decode
\EndFor
\EndProcedure
\end{algorithmic}
\end{algorithm}

In \cite{maddah2013decentralized}, Maddah-Ali and Niesen proved that for any $K,N\in\mathbb{N}^+$, $M\in(0,N]$,
\begin{align}
\frac{R_D(M)}{R_C(M)}\leq 12\notag
\end{align}
Specifically, they also showed via numerical experiment that
\begin{eqnarray*}\label{Eqn_num_bound}
\frac{R_D(M)}{R_C(M)}\leq 1.6
\end{eqnarray*}

\section{Main Results}\label{sec_results}

In this section,  we first introduce the primary result of this paper in Theorem \ref{thm_main}.

\begin{theorem}\label{thm_main}  For $N\in\mathbb{N}^+$ files and $K\ge 2\in\mathbb{N}^+$ users each with a cache of size $M\in(0,N]$, the ratio between the rates achieved by the decentralized coded caching scheme in Algorithm \ref{alg1} and the centralized coded caching scheme in Algorithm \ref{alg2} can be bounded by:
\begin{align}
1\leq\frac{R_D(M)}{R_C(M)}\leq 1.5\label{result1}
\end{align}
Moreover, for any fixed $N\in\mathbb{N}^+$ files and cache size $M\in(0,N]$,
\begin{align}
\lim_{K\rightarrow\infty}\frac{R_D(M)}{R_C(M)}=1\label{result2}
\end{align}

\end{theorem}

Firstly, the lower bound in \eqref{result1} indicates that for any $K,N\in\mathbb{N}^+$, $M\in(0,N]$,
\begin{align}
R_D(M)\geq R_C(M)\notag
\end{align}
Naturally,   since all users' caches are coordinated by the central server jointly, the centralized coded caching scheme can always outperform the decentralized coded caching scheme. Though this fact is intuitively correct, the theoretical justification is not trivial.

Secondly, the upper bound of 1.5 in \eqref{result1} provides a tighter upper estimate of 12 in \cite{maddah2013decentralized}.  Actually, the upper bound is even tighter than the numerical experimental bound of $1.6$ in \cite{maddah2013decentralized}. Indeed, this result indicates that the performance of decentralized scheme is very close to that of the centralized scheme for any system parameters. Furthermore, it is shown in Section \ref{sec_proof} that
\begin{itemize}
\item The lower bound is achieved whenever $\lceil KM/N\rceil<K/N-1$ (see \eqref{eq_ltight});
\item The upper bound is achieved when $K=2,N\geq2$ and $M=N/2$ (see \eqref{eq_rtight});
\end{itemize}
Thus, both the lower bound and the upper bound in \eqref{result1} are tight.

Finally, \eqref{result2} depicts the case of large number of users. It is an essentially surprising result, since it says that the rate of the decentralized coded caching scheme approaches the rate of the centralized coded caching scheme, for any fixed cache size $M$ and number of files $N$. In other words, for a system with large number of users, decentralized coded caching scheme is able to achieve almost the same rate with the centralized coded caching. Thus in cases  where the centralized coded caching scheme can not be implemented, it is reasonable to be replaced by the decentralized counterpart.

\section{Proofs of Theorem \ref{thm_main}}\label{sec_proof}

\subsection{Preliminaries}

%

In order to derive our proof,  it is more convenient to set
\begin{eqnarray}
s=\lceil KM/N \rceil\in\{1,2,\cdots,K\}\notag
\end{eqnarray}  rewrite
\begin{align}
\frac{M}{N}&=\theta\cdot\frac{s-1}{K}+(1-\theta)\cdot \frac{s}{K}=\frac{s-\theta}{K},\notag\\
\theta&=\lceil KM/N \rceil-KM/N\in[0,1),\notag
\end{align}
and re-express $R_C(M)$ given by \eqref{eq_RM} as the following piecewise function

\begin{numcases}{R_C(M)=}
\theta\cdot\frac{ K-s+1}{s}+(1-\theta)\cdot \frac{K-s}{s+1}, \notag \\
~~~~~~~~~~~~~~~~~~~~\mbox{if}~~K/N \le s\le K \label{eq_RCa}\\
                   \theta\cdot N\cdot \frac{K-s+1}{K}+(1-\theta)\cdot\frac{ K-s}{s+1},\notag \\
~~~~~~~~~~~~~~~~~~~~                   \mbox{if}~~K/N-1\leq s< K/N\label{eq_RCb}\\
                  N-M,  &\notag\\
~~~~~~~~~~~~~~~~~~~~                  \mbox{if}~~1\le s< K/N-1&\label{eq_RCc}
\end{numcases}

Further, to  simplify the proof, we define three functions as
\begin{align}
r_C(q,K)&\overset{\triangle}{=}\frac{K(1-q)}{1+Kq}\label{eq_rC}\\
\widetilde{r}_C(q,K)&\overset{\triangle}{=}\theta\cdot r_C\left(\frac{s-1}{K},K\right)+(1-\theta)\cdot r_C\left(\frac{s}{K},K\right)\label{eq_rc}\\
r_D(q,K)&\overset{\triangle}{=}\frac{1-q}{q}\left(1-(1-q)^K\right)\label{eq_rD}
\end{align}
where  $q\in(0,1]$, $K\in\mathbb{N}^+$, $s=\lceil Kq\rceil$, and $\theta=s-Kq$. Obviously,
if setting
 \begin{align}
 q=\frac{M}{N}\notag
 \end{align}
then we get the following relations between $R_C(M)$ and $\widetilde{r}_C(q,K)$,  as well as $R_D(M)$ and $r_D(q,K)$:
\begin{enumerate}
  \item If $K/N\le s\le K$, then
  \begin{align}
  R_C(M)&= \theta \cdot \frac{K-s+1}{s}+(1-\theta)\cdot\frac{ K-s}{s+1}\notag\\
  &=\widetilde{r}_C(q,K)\label{eq_Rr2}
  \end{align}
\item If $K/N-1\leq s<K/N$, then
  \begin{align}
 R_C(M)&= \theta\cdot N\cdot \frac{K-s+1}{K}+(1-\theta)\cdot\frac{ K-s}{s+1}\notag\\
 &<\widetilde{r}_C(q,K)\label{eq_Rr3}
  \end{align}
  \item
\begin{align}
  &R_D(M)\notag\\
  =& \min\{r_D(q,K), N-M\}\label{Eqn_RDm1}\\
  =&\min \left\{\frac{K-s+\theta}{s-\theta}\cdot\left(1-\left(\frac{K-s+\theta}{K}\right)^K\right),\right.\notag\\
   &~~~~~~~~~~~~~~~~~~~~~~~~~~~~~~\left.N\cdot \frac{K-s+\theta}{K}\right\}\label{Eqn_RDm2}
\end{align}
  \end{enumerate}

Finally, we elaborate  the  relations between $r_C(q,K)$, $\widetilde{r}_C(q,K)$, and $r_D(q,K)$  as follows, which
associated with \eqref{eq_Rr2}-\eqref{Eqn_RDm1}, will play a key role in our proof.

\begin{lemma}\label{lem2}
For $\forall~q\in(0,1]$, $\forall~K\in\mathbb{N}^+$,
\begin{align}
r_C(q,K)\leq \widetilde{r}_C(q,K)\leq r_D(q,K)\label{eq_r1}
\end{align}
and moreover, if $K\geq 3$, then
\begin{align}
\frac{r_D(q,K)}{r_C(q,K)}<1.5\label{eq_r2}
\end{align}
Furthermore, for any fixed $q\in (0,1]$,
\begin{align}
\lim_{K\rightarrow\infty}\frac{r_D(q,K)}{r_C(q,K)}=1\label{eq_r3}
\end{align}
\end{lemma}

The proof of Lemma \ref{lem2} is left in Appendix.
\subsection{Proof of The Lower Bound in Theorem \ref{thm_main}}

The proof is divided into  2 cases according to the  expressions of $R_C(M)$ in  \eqref{eq_RCa}-\eqref{eq_RCc}.

\subsubsection*{Case $K/N-1 \le s\le K $}\label{subsec_low1}
 In this case, applying Lemma \ref{lem2} to \eqref{eq_Rr2} and \eqref{eq_Rr3}, we have
 \begin{eqnarray*}
   R_C(M)\le \widetilde{r}_C(q,K)\le r_D(q,K)
 \end{eqnarray*}
 On the other hand,   we always have
  \begin{eqnarray*}
   R_C(M)&\le & \theta \cdot N \cdot \frac{K-s+1}{K}+(1-\theta)\cdot N\cdot\frac{ K-s}{K}\notag\\
         &=& {N\over K }(K-s+\theta)\notag\\
         &=&N\left(1-\frac{s-\theta}{K}\right)\notag\\
         &=&N\left(1-\frac{M}{N}\right)\notag\\
         &=&N-M\notag
 \end{eqnarray*}
 Thus, by \eqref{Eqn_RDm1}, we have
\begin{align}
R_C(M)\leq R_D(M)\notag
\end{align}

\subsubsection*{Case $1\le s<K/N-1$}\label{subsec_low3}
By Lemma \ref{lem2}, we have $r_C(q,K)\le r_D(q,K)$, which gives
\begin{align}
\frac{1}{1+KM/N}<\frac{N}{KM}\left(1-\left(1-\frac{M}{N}\right)^K\right)\notag
\end{align}
for $q=M/N\in (0,1]$ and $K\in \mathbb{N}^+$. Besides, note that
\begin{align}
\frac{N}{K}<\frac{1}{1+s}\leq\frac{1}{1+KM/N}\notag
\end{align}
Then, the above two inequalities imply
\begin{align}
\frac{N}{K}<\frac{N}{KM}\left(1-\left(1-\frac{M}{N}\right)^K\right)\notag
\end{align}
which means $R_D(M)=N-M$ by \eqref{eq_RD}. Whereas $R_C(M)=N-M$ in this case  by \eqref{eq_RCc}, \emph{i.e.},
\begin{align}
R_C(M)=R_D(M)\label{eq_ltight}
\end{align}
Combing these two cases, we derive the lower bound.

\subsection{Proof of The Upper Bound in Theorem \ref{thm_main}}
Hereafter we only need to verify  the upper bound for  $s\geq K/N-1$  since we have shown that $R_C(M)=R_D(M)$ for $1\leq s<K/N-1$ in \eqref{eq_ltight}.

\subsubsection*{Case $K/N\le s\le K$}

\begin{itemize}
  \item If $K=2$, then
  \begin{itemize}
  \item If $N=1$, then $s=2$, as implied by $K/N\le s\le K$. Then by \eqref{eq_RCa} and \eqref{Eqn_RDm2}, we have
    \begin{align}
    R_C(M)=R_D(M)=\frac{\theta}{2}\notag
    \end{align}
    \item If $N\geq2$, then $s=1,2$. By \eqref{eq_RCa} and \eqref{Eqn_RDm2}, we can derive
    \begin{align}
    R_C(M)&=\frac{3\theta+(2-s)s}{s(s+1)}\notag
   \end{align}
and
  \begin{align}
       &R_D(M)\notag\\
       =&\min\left\{\frac{(2-s+\theta)\cdot(4-s+\theta)}{4},N\cdot\frac{2-s+\theta}{2}\right\}\notag\\
    =&\frac{(2-s+\theta)\cdot(4-s+\theta)}{4}\notag
    \end{align}
    Thus,
    \begin{align}
    \frac{R_D(M)}{R_C(M)}&=\frac{s(s+1)}{4}\cdot\frac{(2-s+\theta)\cdot(4-s+\theta)}{3\theta+s(2-s)}\notag\\
    &=\left\{\begin{array}{ll}
           (1+\theta)(3+\theta)/\left(6\theta+2\right),     &\mbox{if}~s=1  \\
               (2+\theta)/2, & \mbox{if}~s=2
             \end{array}
    \right.\notag
    \end{align}
   It is easy to verify that $(1+\theta)(3+\theta)/\left(6\theta+2\right)$ decreases with $\theta$ while $(2+\theta)/2$ increases with $\theta$ on $(0,1)$. Therefore,
   \begin{align}
    \frac{R_D(M)}{R_C(M)}
    &\leq\left\{\begin{array}{ll}
           \left.(1+\theta)(3+\theta)/\left(6\theta+2\right)\right|_{\theta=0},     &\mbox{if}~s=1  \\
               \left.(2+\theta)/2\right|_{\theta=1}, & \mbox{if}~s=2
             \end{array}
    \right.\label{eq_rtight}\\
    &=1.5\notag
    \end{align}
  \end{itemize}
  \item If $K\geq 3$, then by \eqref{eq_Rr2} and \eqref{Eqn_RDm1}
  \begin{align}
  R_D(M)&\leq r_D(q,K)\notag\\
  R_C(M)&=\widetilde{r}_C(q,K)\notag
  \end{align}
 Applying Lemma \ref{lem2}, we have
  \begin{align}
  \frac{R_D(M)}{R_C(M)}\leq\frac{r_D(q,K)}{\widetilde{r}_C(q,K)}\leq\frac{r_D(q,K)}{r_C(q,K)}<1.5\notag
  \end{align}
\end{itemize}

\subsubsection*{Case $K/N-1\leq s<K/N$}
In this case,
\begin{align}
&R_C(M)\notag\\
 =&  \theta\cdot N\cdot \frac{K-s+1}{K}+(1-\theta)\cdot\frac{ K-s}{s+1}\label{eq_Rcd41}\\
 \geq& \theta\cdot\frac{K}{s+1}\cdot\frac{K-s+1}{K}+(1-\theta)\cdot\frac{ K-s}{s+1}\notag\\
=&\frac{K-s+\theta}{s+1}\label{eq_Rcd4}
\end{align}
by \eqref{eq_Rr3}.

\begin{itemize}
  \item If $s=1$, then
\begin{eqnarray}\label{ieq3}
{1\over 2}\le {N\over K}<1
\end{eqnarray}

We can deduce two upper bounds on $R_D(M)/R_C(M)$. On one hand, by \eqref{Eqn_RDm2} and \eqref{eq_Rcd41}, we have
\begin{align}
&\frac{R_D(M)}{R_C(M)}\notag\\
\leq &\frac{N(K-1+\theta)/K}{\theta N+(1-\theta)(K-1)/2}\notag\\
=&\frac{\theta +(1-\theta)(K-1)/K}{\theta+(1-\theta)(K-1)/(2N)}\notag\\
\leq&\frac{\theta +(1-\theta)(K-1)/K}{\theta+(1-\theta)(K-1)/(2(K-1))}\label{ieq5}\\
=&2\cdot\frac{1-(1-\theta)/K}{1+\theta}\label{eq_ineq2}\\
=&\frac{2}{K}\cdot\left(1+\frac{K-2}{1+\theta}\right)\label{eq_ineq}
\end{align}
where in \eqref{ieq5} we use the fact $N\leq K-1$ given by \eqref{ieq3}. On the other hand,
with \eqref{Eqn_RDm2} and \eqref{eq_Rcd4}, we then have
\begin{align}
&\frac{R_D(M)}{R_C(M)}\notag\\
\leq& \frac {(K-1+\theta)/(1-\theta)\cdot (1-((K-1+\theta)/K)^K)}{(K-1+\theta)/2}\notag\\
=&2\cdot \frac{1-(1-(1-\theta)/K)^K}{1-\theta}\label{eq_plus1}\\
=&\frac{2}{K}\cdot\sum_{i=0}^{K-1}\left(\frac{K-1+\theta}{K}\right)^{i}\label{eq_plus2}
\end{align}
Note that, by observing \eqref{eq_ineq2} and \eqref{eq_ineq}, the first upper bound increases with $K$ and decreases with $\theta$.    With the fact that $\left(1-\frac{x}{n}\right)^n$ increases with $n$ for $n>x$, \eqref{eq_plus1} and \eqref{eq_plus2} indicates that the second upper bound  decreases with $K$ and increases with $\theta$. So, we decompose this case into five subcases:
\begin{itemize}
  \item If $2\leq K\leq 3$, with \eqref{eq_ineq}, we have
 \begin{align}
\frac{R_D(M)}{R_C(M)}&\leq\frac{2}{3}\cdot\left(1+\frac{1}{1+\theta}\right)\notag\\
&\leq\frac{4}{3}\notag\\
&< 1.5\notag
\end{align}
\item If $4\leq K\leq 8$, $\theta>6/25$, still with  \eqref{eq_ineq},
\begin{align}
\frac{R_D(M)}{R_C(M)}&<\frac{2}{8}\cdot\left(1+\frac{6}{1+6/25}\right)\notag\\
&=\frac{181}{124}\notag\\
&<1.5\notag
\end{align}
\item If $4\leq K\leq 8$, $\theta\leq 6/25$, with \eqref{eq_plus2}, we have
\begin{align}
\frac{R_D(M)}{R_C(M)}&< \frac{2}{4}\cdot\sum_{i=0}^{3}\left(\frac{3+6/25}{4}\right)^i\notag\\
&\approx 1.4988\notag\\
&<1.5\notag
\end{align}

\item If $K\geq9$, $\theta>1/3$, with \eqref{eq_ineq2},
\begin{align}
\frac{R_D(M)}{R_C(M)}&< \frac{2}{1+\theta}\notag\\
&<\frac{2}{1+1/3}\notag\\
&=1.5\notag
\end{align}

\item If $K\geq 9$, $\theta\leq1/3$, with \eqref{eq_plus2},
    \begin{align}
    \frac{R_D(M)}{R_C(M)}&\leq\frac{2}{9}\sum_{i=0}^8\left(\frac{8+1/3}{9}\right)^i\notag\\
    &\approx 1.4993\notag\\
    &< 1.5\notag
    \end{align}

\end{itemize}

\item If $s\geq 2$, then by \eqref{Eqn_RDm2} and \eqref{eq_Rcd4}, we have
\begin{align}
\frac{R_D(M)}{R_C(M)}&\le \frac{N\cdot (K-s+\theta)/K}{(K-s+\theta)/(s+1)}\notag\\
&= \frac{N}{K}\cdot{(s+1)}\notag\\
&< \frac{s+1}{s}\notag\\
&\leq 1.5\notag
\end{align}
\end{itemize}

Collecting all the above cases, we deduce the desired upper bound.

\subsection{Proof of The Limit in Theorem \ref{thm_main}}\label{sec_proof2}

In this subsection, we prove \eqref{result2}. For any fixed $N\in\mathbb{N}^+$, $M\in(0,N]$,
\begin{itemize}
  \item If $M\geq 1$, since $s=\lceil KM/N\rceil\geq K/N$, then $R_C(M)$ is given by \eqref{eq_Rr2}.  It  is easy to check that $R_D(M)$ is evaluated by the first item in \eqref{eq_RD}, thus $R_D(M)=r_D(q,K)$ by \eqref{Eqn_RDm1}. Therefore,
      \begin{align}
      \frac{R_D(M)}{R_C(M)}&=\frac{r_D(q,K)}{\widetilde{r}_C(q,K)}\notag
      \end{align}
According to \eqref{eq_r1}, we have
      \begin{align}
      1 \le \frac{r_D(q,K)}{\widetilde{r}_C(q,K)} \le \frac{r_D(q,K)}{r_C(q,K)}\notag
      \end{align}
The limit then follows from \eqref{eq_r3}.
  \item If $M<1$, we use the expressions of  $R_C(M)$ and $R_D(M)$ in \eqref{eq_RC} and \eqref{eq_RD}, \emph{i.e.},
  \begin{align}
  R_C(M)&=\left(1-\frac{M}{N}\right)\cdot\min\left\{\frac{K}{1+KM/N},N\right\}\notag\\
  R_D(M)&=\notag\\
  &\left(1-\frac{M}{N}\right)\cdot\min\left\{\frac{N}{M}\left(1-\left(1-\frac{M}{N}\right)^K\right),N\right\}\notag
  \end{align}
  Note that as $K\rightarrow\infty$, the first items in the braces of the above expressions approaches $N/M>N$ for $M<1$, we then conclude that for sufficiently large $K$,
  \begin{align}
  R_D(M)=R_C(M)=N\left(1-\frac{M}{N}\right)\notag
  \end{align}
\end{itemize}
Therefore, the limit is proved.
\section{Conclusions}\label{sec_conclusion}

In this paper, we investigated the ratio between the rates of the decentralized and centralized coded caching schemes proposed by Maddah-Ali and Niesen. We proved  that for any system parameters, the ratio is between $1$ and $1.5$. This verified two facts: Firstly, the centralized scheme always outperforms the decentralized scheme due to the central coordination of the server. Secondly, the rate of the decentralized coded caching scheme is within a constant multiplicative gap  $1.5$ of the rate of the centralized coded caching scheme, which is  tighter than the claimed numerical upper bound $1.6$  in \cite{maddah2013decentralized}. Both the upper bound and the lower bound  are tight since they can be achieved in some cases.
Finally, we showed that when the number of users $K$ goes to infinity, the ratio  always approaches $1$. This suggests that in the systems with large number of users, even without the centralized coordination from the central server, the decentralized coded caching scheme is able to achieve almost the same performance with its centralized counterpart.   Therefore, the decentralized coded caching scheme is a reasonable alternative solution for the centralized coded caching scheme in scenarios that the centralized coordination is unavailable.


\appendix


\subsection{Proof of Lemma \ref{lem2}}

\subsubsection*{Proof of \eqref{eq_r1}}
We prove that for $\forall~q\in (0,1]$, $K\in\mathbb{N}^+$,
\begin{align}
r_C(q,K)\leq \widetilde{r}_C(q,K)\leq r_D(q,K)\notag
\end{align}

Note that $r_C(q,K)$ is convex on $q\in(0,1)$. Moreover, we have
\begin{align}
q&= {s-\theta\over K}\label{eqn_q}\\
&=\theta {s-1\over K}+(1-\theta) {s\over K}\notag
\end{align}
since $s=\lceil Kq\rceil$ and $\theta=s-Kq$.
Then, the first inequality follows from the well-known Jensen's inequality.

For the second inequality, by substituting \eqref{eqn_q} into \eqref{eq_rc} and \eqref{eq_rD} respectively, it is therefore
sufficient to prove
\begin{align}
&\theta\cdot\frac{ K-s+1}{s}+(1-\theta)\cdot \frac{K-s}{s+1}\notag\\
\leq& \frac{1-(s-\theta)/K}{(s-\theta)/K}\left(1-\left(1-\frac{s-\theta}{K}\right)^K\right)\notag\\
\Leftrightarrow\qquad&\frac{s-\theta}{K}\left(\theta\frac{K+1}{s(s+1)}+\frac{K+1}{s+1}-1\right)\notag\\
\leq&\left(1-\frac{s-\theta}{K}\right)\cdot\left(1-\left(1-\frac{s-\theta}{K}\right)^K\right)\notag
\end{align}
for all $K\in\mathbb{N}^+,~s\in\{1,2,\cdots,K\}$ and $\theta\in[0,1)$

Define
\begin{align}
&f(\theta,n,s)\notag\\
=&\left(1-\frac{s-\theta}{n}\right)\cdot\left(1-\left(1-\frac{s-\theta}{n}\right)^n\right)\notag\\
&-\frac{s-\theta}{n}\left(\theta\frac{n+1}{s(s+1)}+\frac{n+1}{s+1}-1\right)\notag\\
=&\frac{n-s}{n(s+1)}+\frac{n+1}{ns(s+1)}\theta^2-\left(1-\frac{s-\theta}{n}\right)^{n+1}\label{eq_f7}
\end{align}

In what follows, we only need to verify $f(\theta,n,s)\geq0\label{eq_f0}$ for any $\theta,n$ and $s$,  \emph{s.t.}, $\theta\in[0,1)$, $n,s\in\mathbb{N}^+$, $n\geq s$.

\begin{itemize}

 \item If $s=n\in\mathbb{N}^+$,
  \begin{align}
  f(\theta,n,n)&=\frac{\theta^2}{n^2}-\left(\frac{\theta}{n}\right)^{n+1}\notag\\
  &=\frac{\theta^2}{n^2}\left(1-\left(\frac{\theta}{n}\right)^{n-1}\right)\notag\\
  &\geq0\notag
  \end{align}

  \item If $s=1$ and $n>s$, then
  \begin{align}
  f(\theta,n,1)&=\frac{n-1}{2n}+\frac{n+1}{2n}\theta^2-\left(\frac{n-1+\theta}{n}\right)^{n+1}\notag\\
  \frac{\partial f(\theta,n,1)}{\partial \theta}&=\frac{n+1}{n}\left(\theta-\left(\frac{n-1+\theta}{n}\right)^n\right)\label{eq_f2}
  \end{align}
and
 \begin{align}
   \frac{\partial^2 f(\theta,n,1)}{\partial \theta^2}&=\frac{n+1}{n}\left(1-\left(\frac{n-1+\theta}{n}\right)^{n-1}\right)> 0,\notag\\
   &~~~~~~~~~~~~~~~~~~~~~~~~~~~\forall~\theta\in(0,1)\label{eq_f3}
  \end{align}
\eqref{eq_f3} tells us that  $\eqref{eq_f2}$ is increasing on $\theta\in(0,1)$. Therefore, it follows  that
  \begin{align}
    \frac{\partial f(\theta,n,1)}{\partial \theta}<  \left.\frac{\partial f(\theta,n,1)}{\partial \theta}\right|_{\theta=1}=0,~\forall~\theta\in(0,1)\notag
  \end{align}
  which indicates that $f(\theta,n,1)$ is decreasing on $\theta\in[0,1)$, thus
  \begin{align}
  f(\theta,n,1)\geq f(1,n,1)=0,~\forall ~\theta\in[0,1)\notag
  \end{align}

  \item If $s\geq 2$ and $n>s$, we begin with a lower bound for $f(\theta,n,s)$. Note that $(1-x)^{n+1}$ is convex on $x\in(0,1)$. Thus, using Jensen's inequality we have
          \begin{align}
          &\left(1-\frac{s-\theta}{n}\right)^{n+1}\notag\\
          =& \left(1-\theta\frac{s-1}{n}-(1-\theta)\frac{s}{n}\right)^{n+1}\notag\\
          \leq& \theta\left(1-\frac{s-1}{n}\right)^{n+1}+(1-\theta)\left(1-\frac{s}{n}\right)^{n+1}\notag
          \end{align}
          Then by \eqref{eq_f7},
           \begin{align}
      &f(\theta,n,s)\notag\\
      \geq&\frac{n-s}{n(s+1)}+\frac{n+1}{ns(s+1)}\theta^2\notag\\
      &-\theta\left(1-\frac{s-1}{n}\right)^{n+1}-(1-\theta)\left(1-\frac{s}{n}\right)^{n+1}\notag\\
      =&\frac{n-s}{n(s+1)}-\left(1-\frac{s}{n}\right)^{n+1}+\frac{n+1}{ns(s+1)}\notag\\
      &\cdot\left(\theta^2-\theta\frac{ns(s+1)}{n+1}\left(\left(1-\frac{s-1}{n}\right)^{n+1}-\left(1-\frac{s}{n}\right)^{n+1}\right)\right)\notag
           \end{align}
      When $\theta=ns(s+1)/(2(n+1))\cdot\left((1-(s-1)/n)^{n+1}-(1-s/n)^{n+1}\right)$, the righthand side achieves its minimal value, which is defined as
      \begin{align}
      &g(n,s)\notag\\
      \overset{\triangle}{=}&\frac{n-s}{n(s+1)}-\left(1-\frac{s}{n}\right)^{n+1}-\frac{1}{4}\cdot\frac{ns(s+1)}{n+1}\notag\\
      &\cdot\left(\left(1-\frac{s-1}{n}\right)^{n+1}-\left(1-\frac{s}{n}\right)^{n+1}\right)^2\notag\\
      =&\frac{1}{s+1}-\frac{s}{n(s+1)}-\left(1-\frac{s}{n}\right)^{n+1}-\frac{s(s+1)}{4}\cdot\frac{n}{n+1}\notag\\
      &\cdot\left(1-\frac{s-1}{n}\right)^{2n+2}\cdot\left(1-\left(1-\frac{1}{n-s+1}\right)^{n+1}\right)^2\notag
      \end{align}

   Note the fact that both the items $(1-s/n)^{n+1}$ and $\left(1-(s-1)/n\right)^{2n+2}$ strictly increases with $n> s$ and
   \begin{align}
   \lim_{n\rightarrow\infty}\left(1-\frac{s}{n}\right)^{n+1}&=e^{-s}\notag\\
   \lim_{n\rightarrow\infty}\left(1-\frac{s-1}{n}\right)^{2n+2}&=e^{-2s+2}\notag
   \end{align}
   which implies $(1-s/n)^{n+1}<e^{-s}$ and $\left(1-(s-1)/n\right)^{2n+2}<e^{-2s+2}$.
   Therefore, we have
   \begin{align}
  & g(n,s)\notag\\
   \geq&\frac{1}{s+1}-\frac{s}{n(s+1)}-e^{-s}-\frac{e^2}{4}\cdot\frac{s(s+1)}{e^{2s}}\notag\\
   &\cdot\left(1-\left(1-\frac{1}{n-s+1}\right)^{n+1}\right)^2\notag\\
   \overset{\triangle}{=}&h(n,s)\notag
   \end{align}

 Up to now, we have derived
   \begin{align}
   f(\theta,n,s)\geq g(n,s)\geq h(n,s)\quad \forall~\theta\in [0,1)\label{eq_fgh}
   \end{align}

   Note that $h(n,s)$ increases with $n$ since the positive number $\left(1-1/(n-s+1)\right)^{n+1}$ increases with $n$. Therefore,

          \begin{itemize}
            \item If $s\geq4$,  it is easy to check that $e^{2s}>4s(s+1)^3$ and $e^s>2(s+1)^2$ for $s\geq 4$. Consequently, we have
        \begin{align}
        &h(n,s)\notag\\
        \ge& h(s+1,s)\notag\\
        >&\frac{1}{s+1}-\frac{s}{(s+1)^2}-e^{-s}-\frac{e^2}{4}\cdot\frac{s(s+1)}{e^{2s}}\notag\\
        >&\frac{1}{s+1}-\frac{s}{(s+1)^2}-\frac{1}{2(s+1)^2}-\frac{e^2}{16}\cdot\frac{1}{(s+1)^2}\notag\\
        =&\left(\frac{1}{2}-\frac{e^2}{16}\right)\cdot\frac{1}{(s+1)^2}\notag\\
        >&0\label{eqn_ks1}
        \end{align}
        where in  \eqref{eqn_ks1} we make use of the fact $e^2<8$.
            \item If $s=3$,  $n\geq 5$, we have
           \begin{align}
            h(n,3)&\geq h(5,3)\approx 0.0045>0\notag
            \end{align}
            \item If $s=2$, $n\geq 8$, we have
            \begin{align}
            h(n,2)&\geq h(8,2)\approx 0.0004>0\notag
            \end{align}
      \item For other cases, \emph{i.e}, $s=3, n=4$ or $s=2, n=3,4,5,6,7$, we  compute $g(n,s)$ directly
      \begin{align}
      g(4,3)\approx 0.0593>0,\quad g(3,2)\approx 0.0602>0\notag\\ g(4,2)\approx 0.0845>0,\quad g(5,2)\approx0.0953>0\notag\\
      g(6,2)\approx0.1012>0,\quad g(7,2)\approx 0.1047>0\notag
      \end{align}
      \end{itemize}
 Therefore, we conclude  $f(\theta,n,s)>0$ for $\theta\in[0,1)$ and $2\le s<n$, by \eqref{eq_fgh}.

\end{itemize}

\subsubsection*{Proof of \eqref{eq_r2}}
We prove that
\begin{align}
\frac{r_D(q,K)}{r_C(q,K)}<1.5 \notag
\end{align}
for all $K\ge 3\in\mathbb{N}^+$ and $q\in(0,1]$.

Note from \eqref{eq_rC} and \eqref{eq_rD},
\begin{align}
\frac{r_D(q,K)}{r_C(q,K)}
&=\frac{1+Kq}{Kq}\left(1-(1-q)^K\right)\label{eq_rDC}\\
&= \frac{1+x}{x}\left(1-\left(1-\frac{x}{K}\right)^K\right)\notag
\end{align}
where $x\overset{\triangle}{=}Kq$.  Therefore, it suffices to show
\begin{align}
l_n(x)\overset{\triangle}{=}\frac{1+x}{x}\left(1-\left(1-\frac{x}{n}\right)^n\right)<1.5 \notag
\end{align}
for any $x\in(0,n]$, where $n\geq3$ and $n\in\mathbb{N}^+$.

\begin{itemize}
  \item If $x\in(0,3]$,
  \begin{align}
l_n(x)&\leq\sup_{z\in(0,3]} l_n(z)\notag\\
&\leq \sup_{z\in(0,3]}l_3(z)\label{eq_h5}\\
&=\sup_{z\in(0,3]}\frac{z^3-8z^2+18z+27}{27}\notag\\
&=\left.\frac{z^3-8z^2+18z+27}{27}\right|_{z=\frac{8-\sqrt{10}}{3}}\notag\\
&=\frac{1001+20\sqrt{10}}{729}\notag\\
&\approx 1.4599\notag\\
&<1.5\notag
\end{align}
where in \eqref{eq_h5}  we use  the fact that for any fixed $z\in(0,3]$, $\left(1-z/n\right)^n$ increases with $n$.
  \item If $3<x\leq n$, then
  \begin{align}
  l_n(x)&\leq \frac{1+x}{x}\notag\\
  &<\frac{1+3}{3}\notag\\
 &<1.5\notag
  \end{align}
\end{itemize}

\subsubsection*{Proof of \eqref{eq_r3}} The limit \eqref{eq_r3}  is clear from the expression \eqref{eq_rDC}.

\ifCLASSOPTIONcaptionsoff
\fi
\bibliographystyle{IEEEtran}

\end{document}